\documentclass[a4paper]{jpconf}
\usepackage{graphicx}
\usepackage{caption}
\usepackage{url}

\usepackage[version=3]{mhchem}

\usepackage{amsmath}
\usepackage{amssymb}
\usepackage[usenames,dvipsnames]{xcolor}
\usepackage{soul}

\begin{document}


\title{Development and tests of a new prototype detector for the XAFS beamline at Elettra Synchrotron in Trieste}

\author{S Fabiani$^1$, M Ahangarianabhari$^2$,G Baldazzi$^{3}$, P Bellutti$^4$, G Bertuccio$^{2,5}$, M Bruschi$^{6}$, J Bufon$^{7,1,8}$, S Carrato$^8$, A Castoldi$^{5,9}$, G Cautero$^{7,1}$, S Ciano$^{1}$,A Cicuttin$^{10}$ , M L Crespo$^{10}$, M Dos Santos$^{10}$ , M Gandola$^{2}$, G Giacomini$^4$, D Giuressi$^7$, C Guazzoni$^{5, 9}$, R H Menk$^{7,1}$,J Niemela$^{10}$, L Olivi$^{7,1}$, A Picciotto$^4$, C Piemonte$^4$, I Rashevskaya$^{11}$, A Rachevski$^1$, L P Rignanese$^{3}$, A Sbrizzi$^{6}$, S Schillani$^{7}$, A Vacchi$^{1, 12}$, V Villaverde Garcia$^{10}$, G Zampa$^1$, N Zampa$^1$, N Zorzi$^4$} 

\address{$^1$ Istituto Nazionale di Fisica Nucleare, Section of Trieste, Padriciano 99, I-34149, Trieste, Italy}
\address{$^2$ Politecnico di Milano, Via Anzani 42, I-22100, Como, Italy}
\address{$^{3}$ Universit\'a di Bologna, Via Zamboni 33, I-40126, Bologna, Italy}
\address{$^4$ Fondazione Bruno Kessler  -  FBK,  via Sommarive 18, I-38123, Trento,  Italy}
\address{$^5$ INFN Milano, Via Celoria 16, I-20133, Milano, Italy}
\address{$^6$ INFN Sezione di Bologna,  Via Irnerio 46, I-40126, Bologna, Italy}
\address{$^7$ Elettra-Sincrotrone Trieste S.C.p.A., SS14, I-34012, Trieste, Italy}
\address{$^8$ Universit\'a di Trieste, Piazzale Europa 1, I-34128, Trieste, Italy}
\address{$^9$ Politecnico di Milano, Piazza Leonardo da Vinci 32, I-20133, Milano, Italy}
\address{$^{10}$ International Centre for Theoretical Physics - MLAB, Via Beirut 31,I-34014, Trieste, Italy}
\address{$^{11}$ TIFPA-INFN, via Sommarive 14, I-38123, Trento, Italy}
\address{$^{12}$ Universit\'a di Udine, Via delle Scienze 206, I-33100, Udine, Italy}

\ead{sergio.fabiani@ts.infn.it}

\begin{abstract}
The XAFS beamline at Elettra Synchrotron in Trieste combines X-ray absorption spectroscopy and X-ray diffraction to provide chemically specific structural information of materials. It operates in the energy range 2.4-27 keV by using a silicon double reflection Bragg monochromator. The fluorescence measurement is performed in place of the absorption spectroscopy when the sample transparency is too low for transmission measurements or the element to study is too diluted in the sample. We report on the development and on the preliminary tests of a new prototype detector based on Silicon Drift Detectors technology and the SIRIO ultra low noise front-end ASIC. The new system will be able to reduce drastically the time needed to perform fluorescence measurements, while keeping a short dead time and maintaining an adequate energy resolution to perform spectroscopy. The custom-made silicon sensor and the electronics are designed specifically for the beamline requirements.
\end{abstract}

\section{Introduction}
\begin{figure} 
 \begin{center}
\begin{tabular}{c}
\includegraphics[scale=0.33]{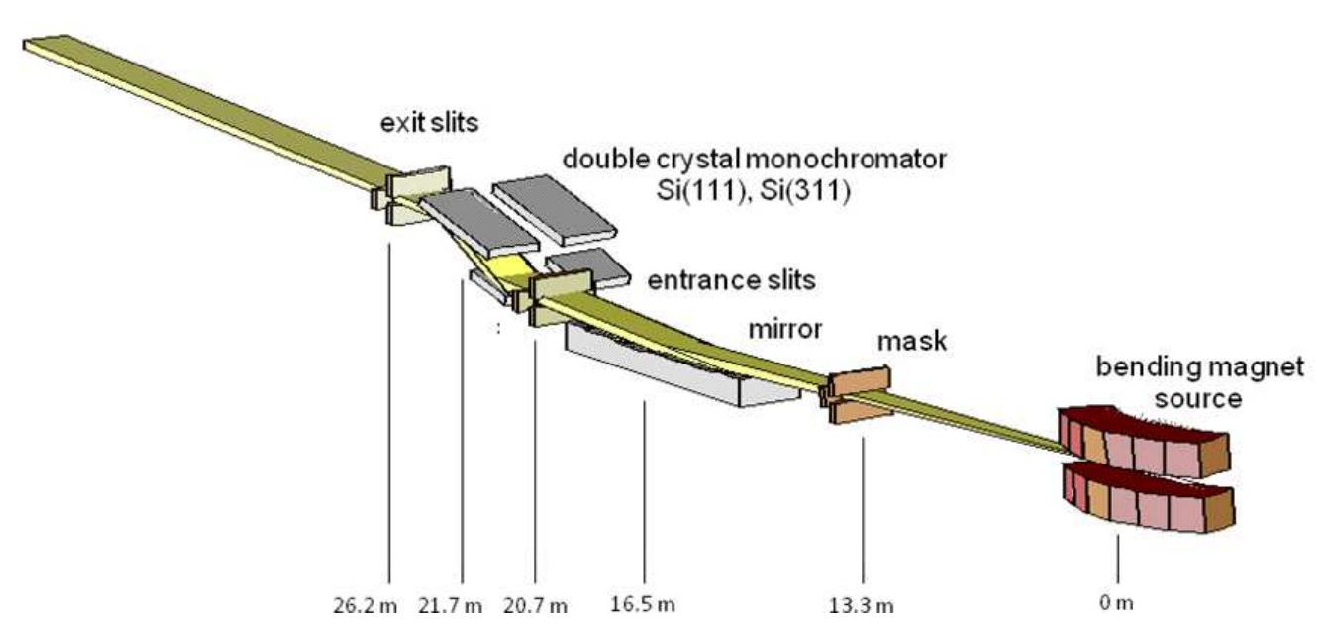}
\end{tabular}
\caption{Scheme of the XAFS beamline. Figure from \cite{DiCicco2009}}\label{fig:Fig1XAFSscheme}
 \end{center}
 \end{figure}

The XAFS (X-ray absorption fine structure) beamline at Elettra Synchrotron in Trieste combines the capability of X-ray absorption spectroscopy (XAS) with X-ray diffraction (XRD) to provide chemically specific structural information. Both techniques allow to fully characterize material structure.
The XAFS spectrometer \cite{DiCicco2009} is located  along the tangential fan of a bending magnet as shown in the sketch of Fig.~\ref{fig:Fig1XAFSscheme}. 
The monochromator (a double flat crystal, double cam Kohzu apparatus) allows the XAFS beamline to operate in the energy range 2.4--27 keV by using an interchangeable pairs of Si(111) and Si(311) crystals on which X-rays undergo two successive Bragg reflections.
Four tungsten alloy slits placed after the monochromator are used to define the beam section impinging on the sample.
 
The beamline is designed for spectroscopic analysis aimed at studying the behaviour of the absorption coefficient as a function of energy in the XAS region (absorption spectroscopy). The energy scan is performed typically in an energy range of 1--2 keV around the energy of interest by operating on the Si double crystal monochromator, so to sample between 1000 and 2000 energy points.

However, the fluorescence analysis is performed in place of the absorption spectroscopy
when the sample is optically thick and does not allow for transmission measurements.
With the present set-up about 5--10 s of integration are needed to have enough statistics for each energy value. Therefore, up to $2\times10^4$ s (about 6 hours) are needed to sample $2000$ energy points.

A sample typically comprises a matrix of various elements in addition to the one of interest. All of them could be excited via fluorescence.
Moreover, radiation scattering can occur on the set-up spreading photon energy across the energy spectrum.
Due to these effects sometimes only a very small fraction of the emitted radiation from the specimen comes from the elements of interest. Therefore, the detector must be able to sustain a high count rate with an adequate energy resolution to perform the measurements.
Currently, a KETEK GmbH AXAS-M Silicon Drift Detector (SDD) is employed for the fluorescence measurements. It has a single 100 mm$^2$ SDD cell (80 mm$^2$ of effective area with collimation) with an FWHM (Full Width Half Maximum) energy resolution of $\sim$170 eV for the Mn K$\alpha$ line at 5.89 keV for a peaking time of 1$\sim \mu s$ at -70$^\circ$C. This set-up cannot sustain a high count rate. This is mainly due to the dead time caused by the monolithic geometry of the Si sensor.

This work reports on the study carried out to design a new fluorescence detector, based on the Silicon Drift Detector technology, able to maintain a short dead time and an adequate energy resolution to perform spectroscopy. This goal will be achieved by manufacturing a new custom-made detector designed specifically for the beamline requirements.

\section{The detector}
The new detector will have a larger area and will be segmented to maintain the pile-up at low level. A sensor will comprise small cells each one constituting an independent read-out channel. Small cells will have a low leakage current, of the order of 10~pA at ambient temperature, allowing to perform measurements with a moderate cooling. Therefore, the cooling system will not be a critical issue even for the large area detector.
Its final version will be manufactured to work both in air and in vacuum.
\begin{figure}[htbp]
\centering
\begin{minipage}[c]{.40\textwidth}
\centering\setlength{\captionmargin}{0pt}%
\includegraphics[width=0.8\textwidth]{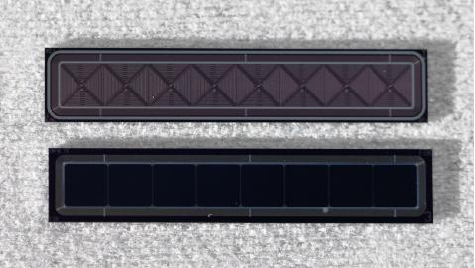}
\caption{SDD linear array comprising 8 squared cells $3\times 3$~mm$^2$. Anode (top) and entrance window (bottom) sides.}\label{fig:SDD}
\end{minipage}%
\hspace{10mm}%
\begin{minipage}[c]{.37\textwidth}
\centering\setlength{\captionmargin}{0pt}%
\includegraphics[width=1.\textwidth]{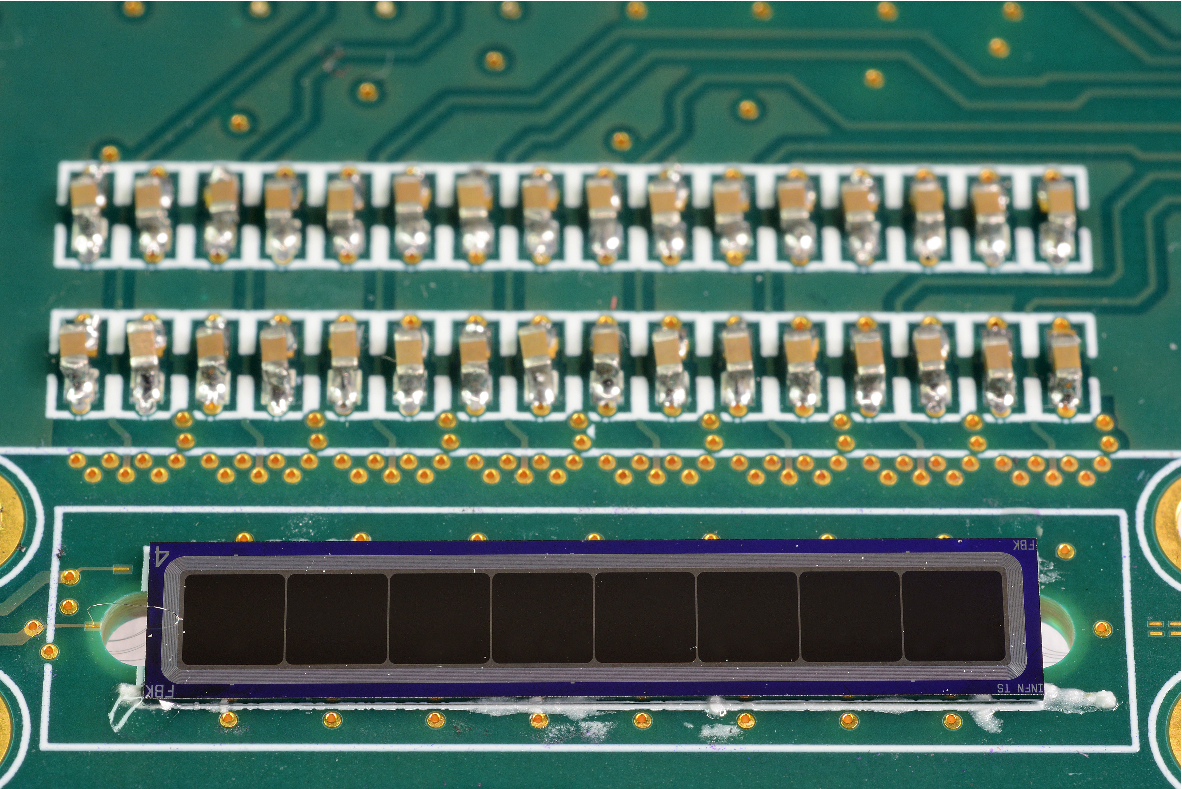}
\caption{The SDD front side mounted on the front-end electronics board.}\label{fig:frontside}
\end{minipage}
\hspace{20mm}%
\begin{minipage}[c]{.37\textwidth}
\centering\setlength{\captionmargin}{0pt}%
\includegraphics[width=1\textwidth]{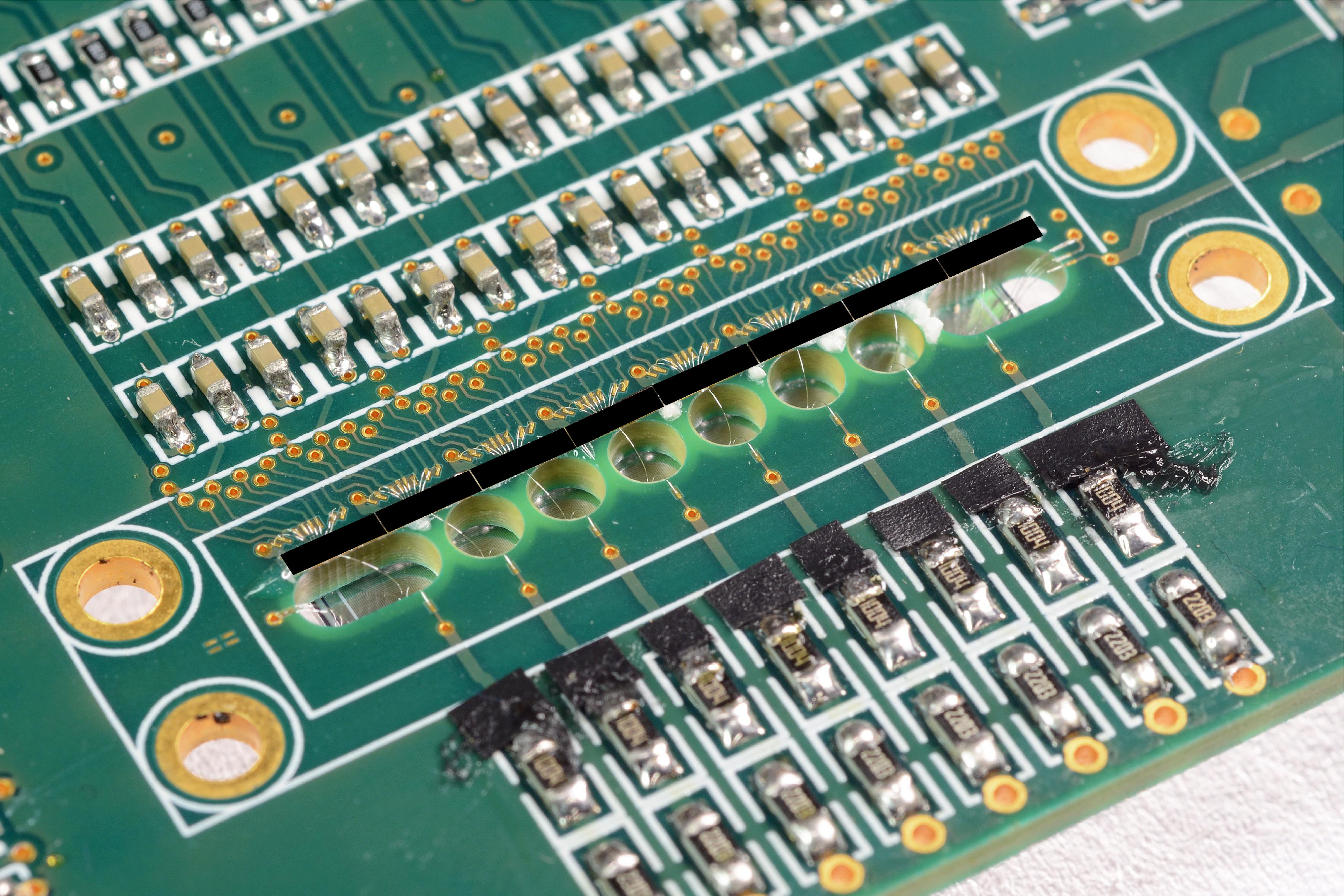}
\caption{Back-end electronics with SDD back side bonded to the preamplifiers.}\label{fig:backside}
\end{minipage}
\hspace{10mm}%
\begin{minipage}[c]{.40\textwidth}
\centering\setlength{\captionmargin}{0pt}%
\includegraphics[width=1\textwidth]{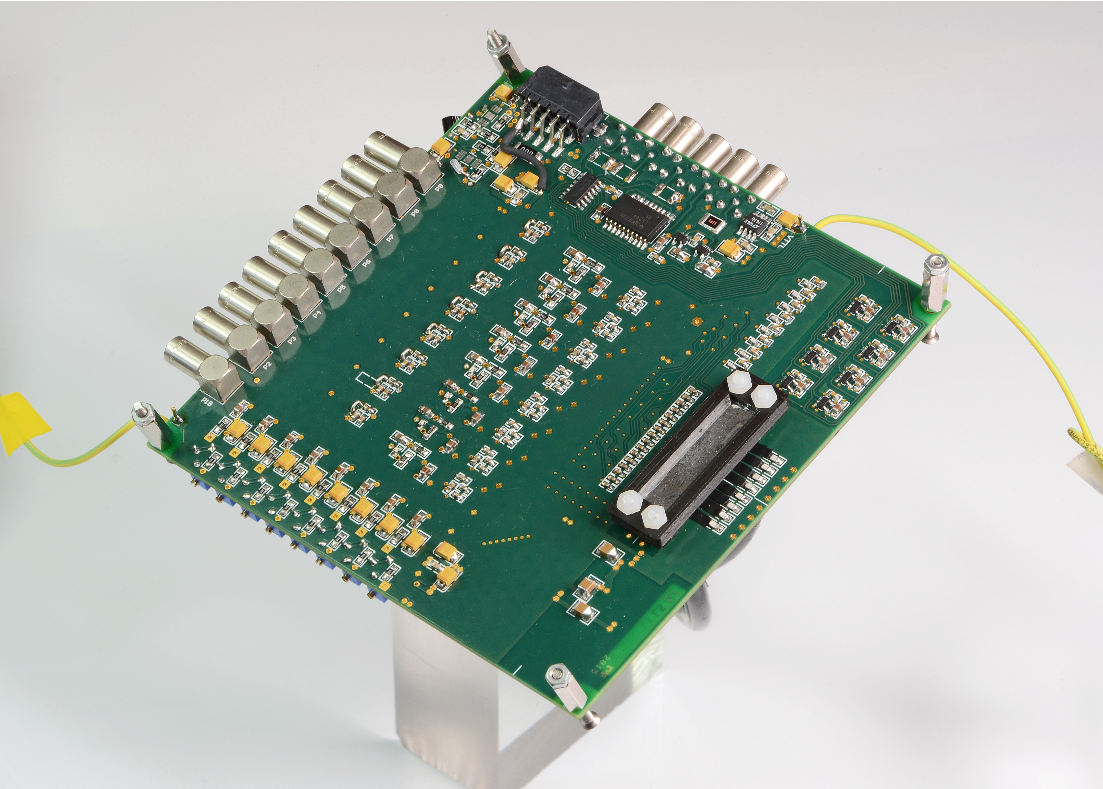}
\caption{Front end electronics board with the plastic box enclosing the SDD.}\label{fig:frontend}
\end{minipage}
\end{figure}

In this work we report on the design activity and tests of the prototype detector given by a single SDD sensor (representative of the final version) coupled to prototypal front-end and back-end electronics boards.
We developed a simulation software based on the GEANT4 toolkit \cite{Geant42010} to optimize the detector design. The simulator reproduces the beamline source energy spectrum and the beam shape out of the last set of four tungsten alloy slits before the sample. To use the detector 
at the highest count rate conditions, that is measuring fluorescence from thick foils of pure elements, we estimated that each detector channel must be able to sustain a maximum data rate of the order of 10$^5$~counts~s$^{-1}$. 
A single SDD sensor module (see Fig.~\ref{fig:SDD}) is designed as an array of 8 squared cells $3\times 3$~mm$^2$. 
The prototype detector comprises a single SDD array (see Fig.~\ref{fig:frontside}), while the final version will be made of $8 \times 8$ matrix for a total collecting area of about 576~mm$^2$, seven times larger than the current detector. 

The SDD sensors of the new detector are obtained from high resistivity n-type silicon wafers 450 $\mu$m thick. They are designed by INFN-Ts and manifactured by FBK.
Their layout is such that the energy of an absorbed X-ray photon is converted into electron-hole pairs and electrons are drifted towards the anodes (small n$^{+}$ pad) on the back side at the center of each cell. The drift cathodes on the back side are arranged as a decreasingly negative biased p$^{+}$ rings.
At the border of each cell the outermost drift cathode is kept at the bias voltage and voltage dividers are integrated separately for each cell to generate potential drops down to the innermost one.
The entrance window is biased separately with respect to the detector back side allowing for an effective charge collection.
Both on the back and the front side, outside the sensitive area of the whole cell array, there are guard cathodes that scale down the bias voltage to ground.
The readout anodes are wire bonded to ultra-low noise SIRIO charge sensitive preamplifiers on the front-end board (see Fig.~\ref{fig:backside}).
The SIRIO preamplifier is designed for SDDs readout in CMOS technology.
It has intrinsic minimum noise levels of 1.2 and 0.9 electrons r.m.s. at +22$^\circ$C and -30$^\circ$C, which correspond to line's FWHM of about 10 eV and 8 eV, respectively, referring to silicon detectors \cite{Bertuccio2014}. 
 This performance represents the current state of the art in low-noise front-end electronics for SDDs and 
the final energy resolution is practically determined by the detector anode dark current and by the 
 parasitic capacitance of the preamplifier-detector connection. An intrinsic noise as low as 8.6 electrons 
 r.m.s., corresponding to 74 eV FWHM on the pulser line, has been demonstrated with a 13~mm$^2$ SDD-SIRIO 
 system operating at room temperature \cite{Bertuccio2015}. The used ASIC is composed by SIRIO and an output amplifier able to drive a 50~$\Omega$ coaxial cable up to 2~m length with a total power consumption of about 10~mW.

The output of the preamplifiers are low-pass filtered to provide anti-aliasing for the following 12~bit 8-channel ADC, capable of encoding  at 40~Msps. The digital data are subsequently treated with a set of suitable digital filters and analysed by an FPGA (Field-Programmable Gate Array), which also handles the near-saturation reset of the preamplifiers. The reset of all the preamplifiers is simultaneous and it is driven by the first one reaching the near-saturation threshold.
The acquired data are transmitted over a TCP~/~IP connection to a PC running a dedicated LabVIEW software.
The back-end FPGA encodes also a shift register hosted on the front-end electronics that allows to disable separately each one of the eight channels by holding the corresponding preamplifier in a reset state.
Since the sensor is sensitive to light, a small plastic box was designed to enclose and keeping it in a dark environment. The X-ray entrance window was realized stacking two foils $25~\mu$m thick of alumined Mylar (see Fig.~\ref{fig:frontend}). 

\section{Tests in laboratory and measurements at the beamline }

\begin{figure} 
\begin{center}
\begin{tabular}{c}
\includegraphics[scale=0.25]{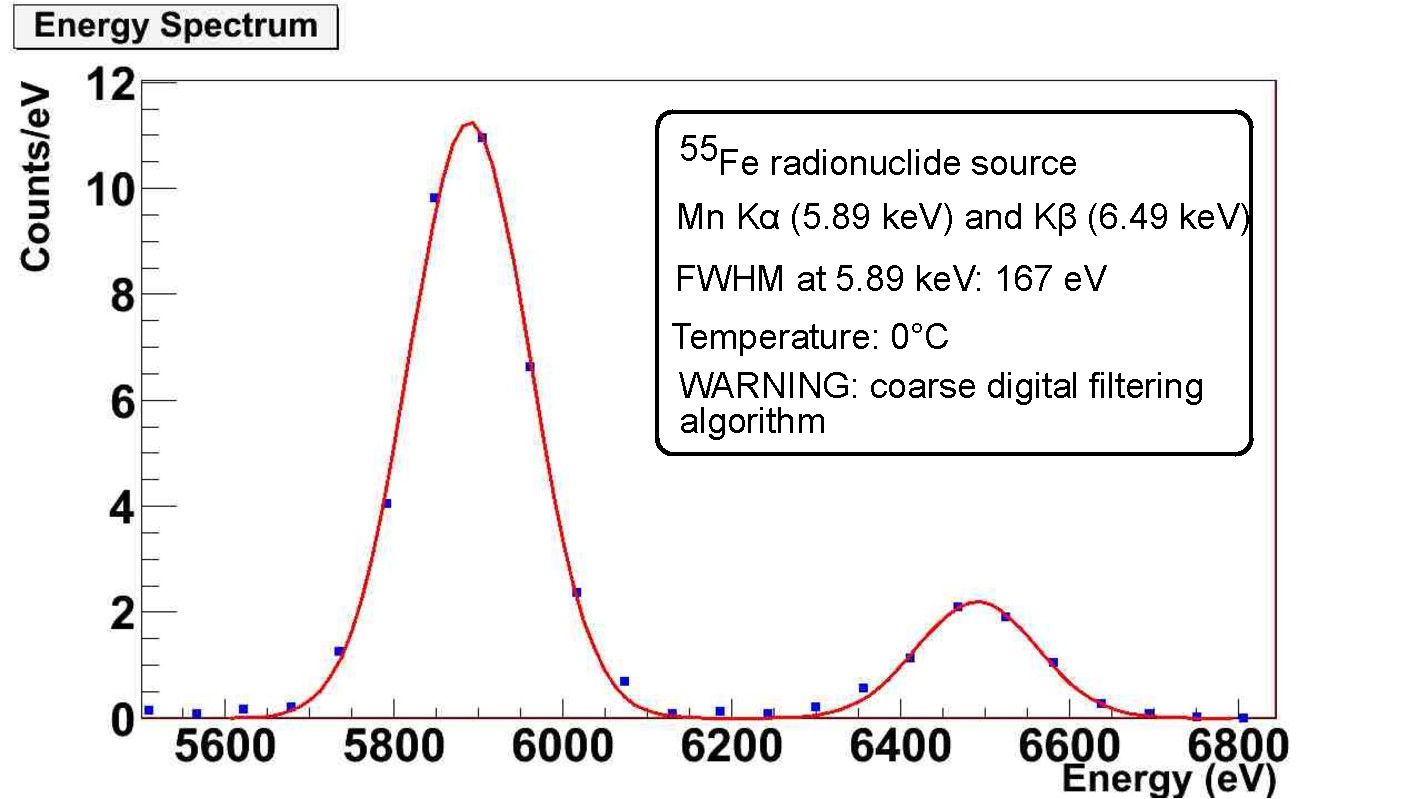}
\end{tabular}
\caption{Energy spectrum of an $^{55}$Fe source at $0^{\circ}$C measured in the climatic chamber.}\label{fig:energyresolution}
\end{center}
\end{figure}
A series of characterization tests were performed on the prototype detector in laboratory at ambient temperature and in climatic chamber down to $0^{\circ}$C. 
The energy resolution was measured by using a coarse off-line algorithm (not yet optimized) for the data analysis. 
In Fig.~\ref{fig:energyresolution} the energy spectrum of an $^{55}$Fe source at $0^{\circ}$C measured in the climatic chamber is shown. The energy resolution of the Mn K$\alpha$ line at 5.89~keV is 167~eV.
 
 \begin{figure}[htbp]
\centering
\begin{minipage}[c]{1\textwidth}
\centering\setlength{\captionmargin}{0pt}%
\includegraphics[scale=0.218]{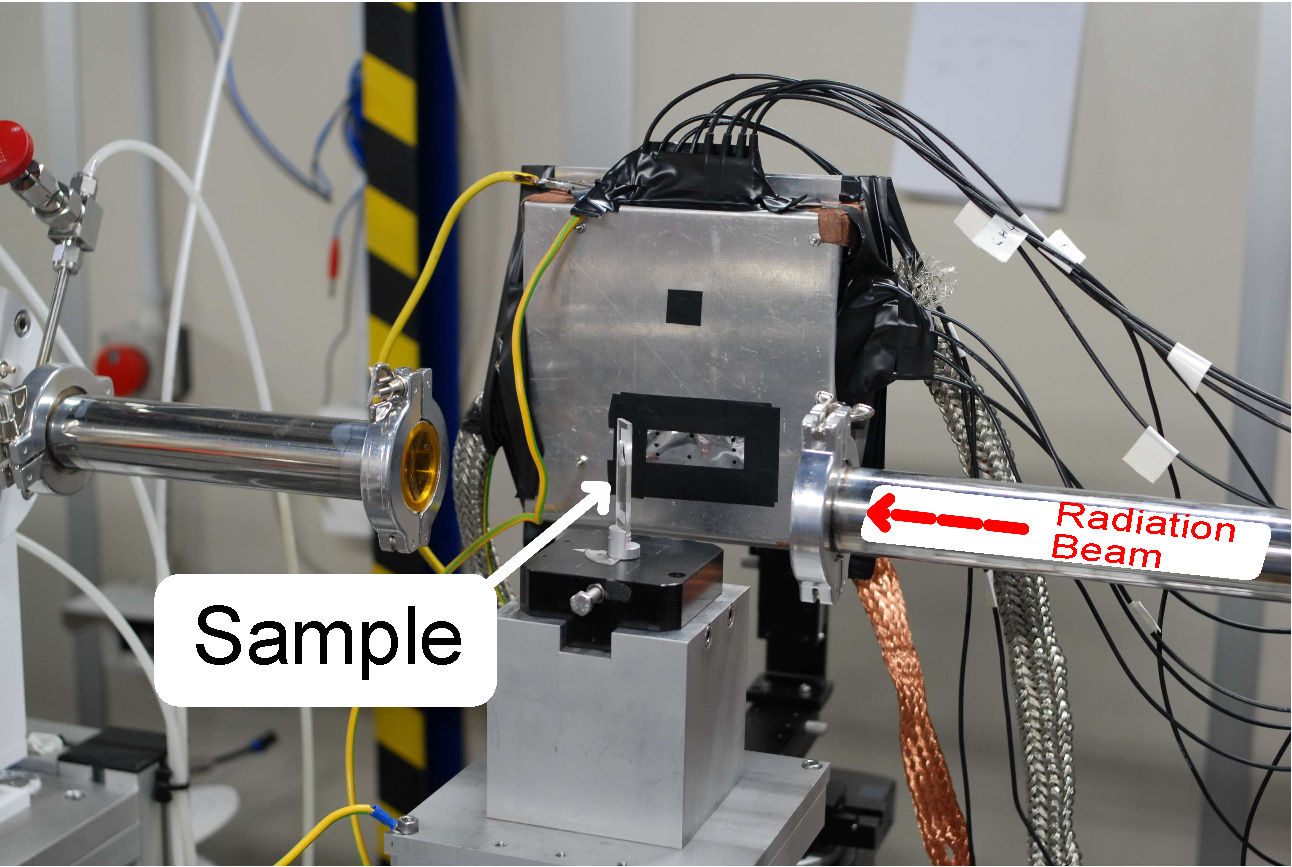}
\includegraphics[scale=0.0528]{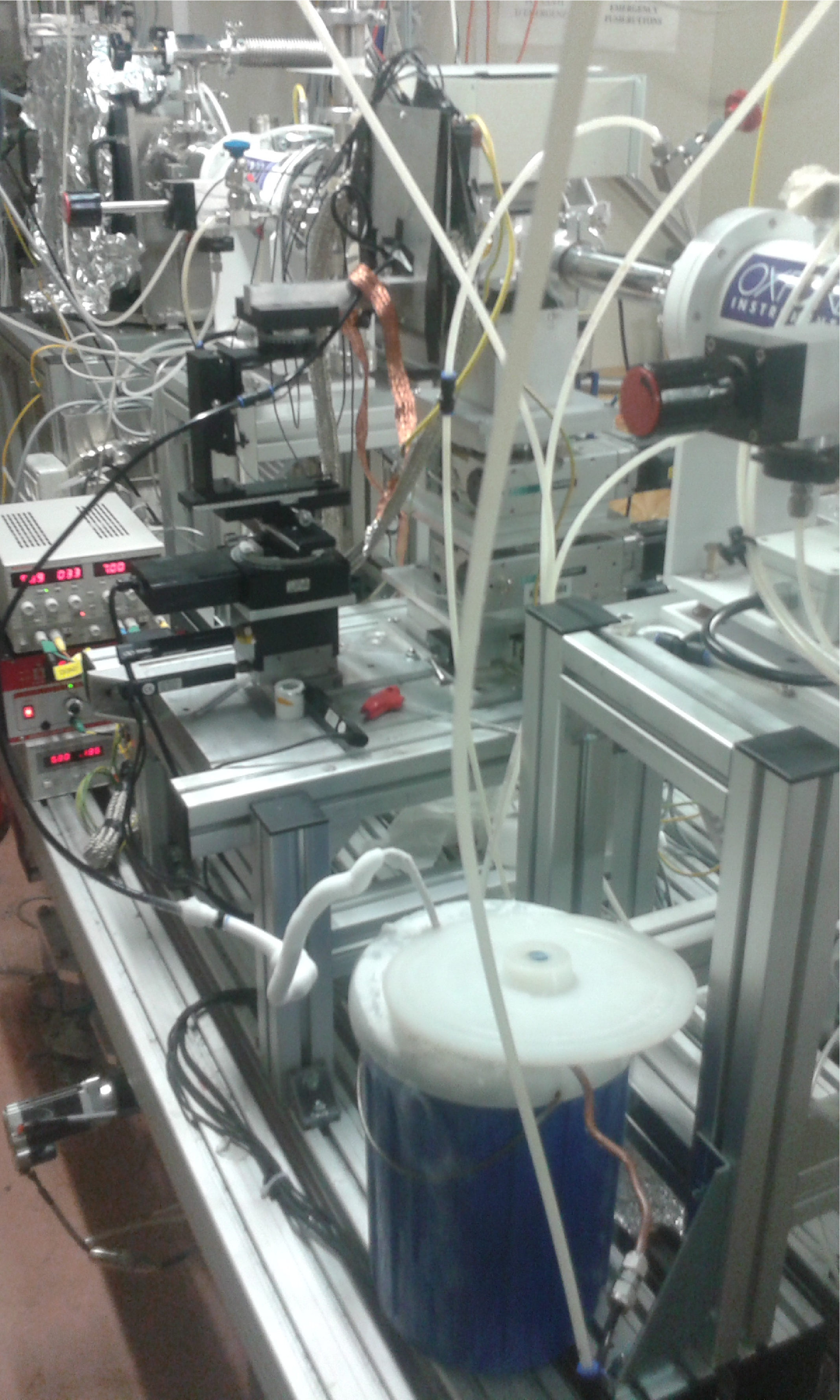}
\caption{On the left: Measurement set-up of the beam test at the XAFS beamline. A foil of a pure element (Mn or Zr) is excited by the radiation beam and emits fluorescence photons. On the right: the dewar filled with liquid nitrogen used to refresh the nitrogen flux that allows the moderate cooling.}\label{fig:beamtestsetup}
\end{minipage}%
\hspace{10mm} \\
\begin{minipage}[c]{1\textwidth}
\centering\setlength{\captionmargin}{0pt}%
\includegraphics[scale=0.22]{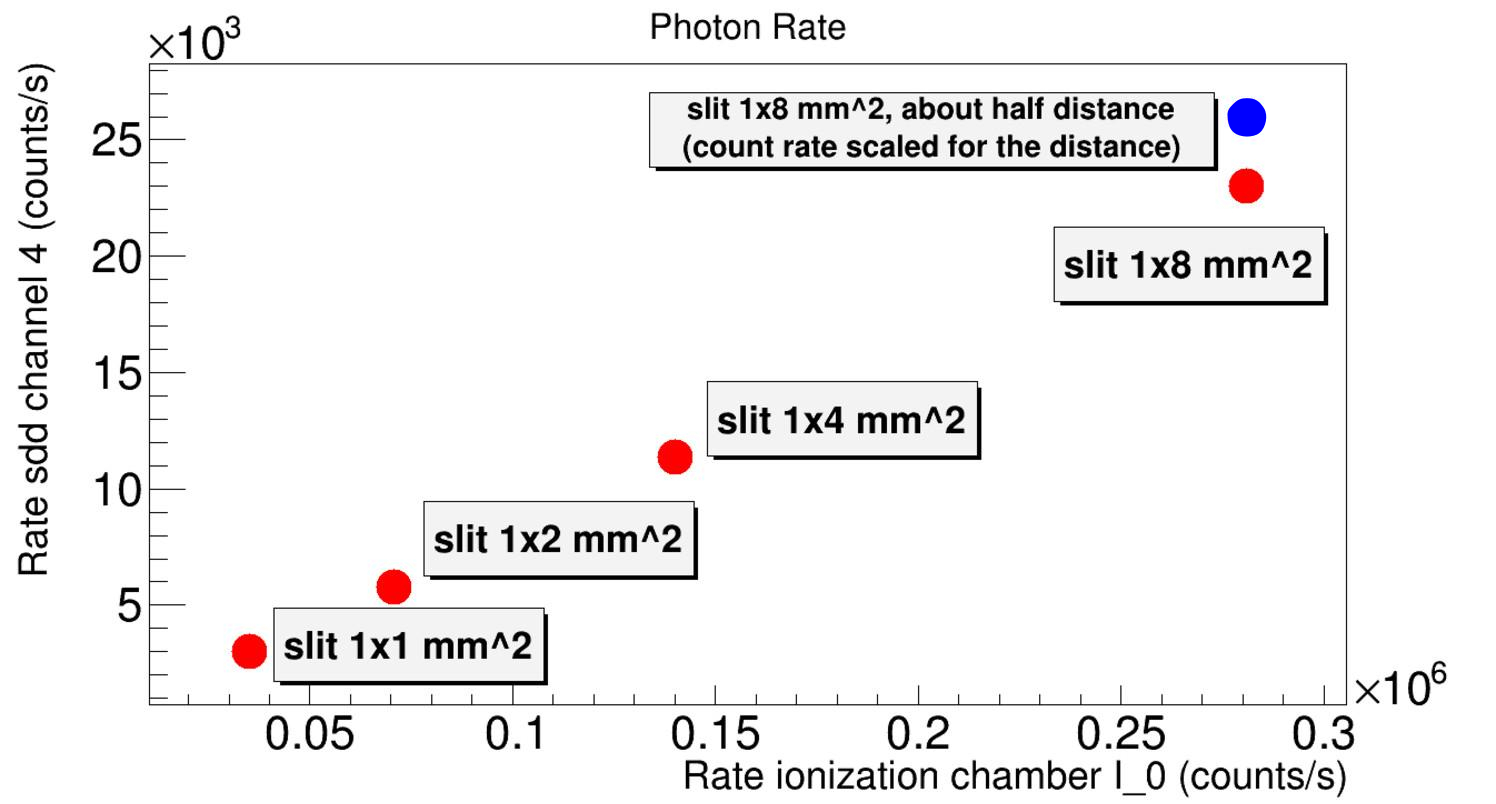}
\caption{Measured count rate on a single SDD cell as a function of the count rate of the first ionization chamber of the XAFS beamline. The highest count rate (blue point) was scaled by a factor 1/4$^{th}$ (see Fig.~\ref{fig:rate2}) to be compared to the other measurements (red points) taken at a larger SDD/sample distance.}\label{fig:rate}
\end{minipage}%
\hspace{10mm} \\ 
\begin{minipage}[c]{1\textwidth}
\centering\setlength{\captionmargin}{0pt}%
\includegraphics[scale=0.25]{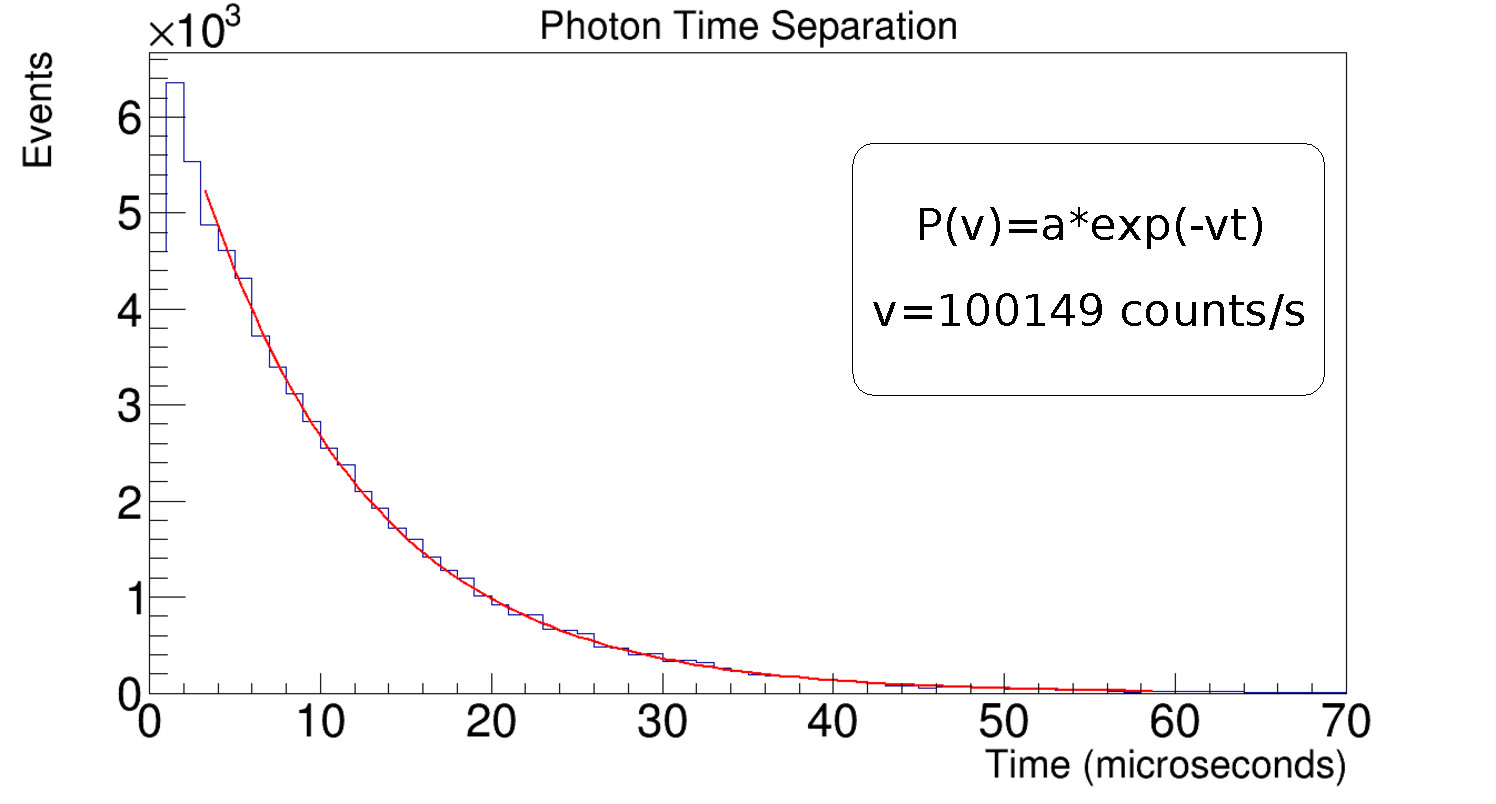}
\caption{Exponential fit of the Poisson distribution of photon arrival time with a slit opening of $1\times 8$~mm$^2$ and 5.5~cm of distance between the sample and the SDD. A count rate of about $10^5$~counts/s was reached without any significant limitation due to pile-up.
}\label{fig:rate2}
\end{minipage}
\end{figure}

A beam test with the prototype detector was carried out at the XAFS beamline of Elettra in the period 16-18 September 2015. 
The measurement set-up is shown in Fig.~\ref{fig:beamtestsetup}. A foil of Mn (K$\alpha$ at 5.89 keV and K$\beta$ 6.49 keV) and then Zr (K$\alpha$ at 15.75 keV and K$\beta$ 17.67 keV) were placed in front of the detector a $45^\circ$ with respect to the beam axis. The front-end board hosting the SDD was inside an aluminum shielding box which entrance window was covered with a $25~\mu$m thick alumined Mylar foil to stop stray visible light entering in the box. The three alumined Mylar foils (two on the entrance window of the plastic box on the front-end electronics board and the third one on the aluminum box), have a total thickness of $75 \mu$m for a transparency of $84.5\%$ at 6~keV. The prototype detector was operated at nearly ambient temperature by fluxing in the plastic box a moderately cooled nitrogen flux. This system allowed to remove the heat produced by electronics components and preamplifiers enclosed in the plastic box.

We tested the detector functionality at different photon fluxes. The photon flux on the sample, and thus the count rate of the detector, depend on the opening of the last set of slits that was increased by steps. Moreover, the count rate was further increased by halving the distance between the sample and the SDD from about 10.8~cm to about 5.5~cm. Using this set-up, we reproduced a very high count rate condition, far away from the working capabilities of the current detector.
In Fig.~\ref{fig:rate} we report the measured count rate with the Mn sample on a single SDD cell as a function of the count rate of the first ionization chamber that is crossed by radiation before impinging on the sample. The red points were obtained placing the detector at 10.8~cm from the sample. The blue one correspond to the count rate measured by placing the detector close to the sample (5.5~cm) and corrected to take into account the different detector to sample distances. Therefore, this count rate value of $100149$~counts/s (see Fig.~\ref{fig:rate2}) was scaled by the ratio of the squares of the two distances (1/4$^{th}$). Since the scaled value is not smaller than the value obtained at 10.8~cm, we conclude that this measurement is not limited by pile-up. 

The beam test was successful and detailed results will be published in a forthcoming article.

\section{Conclusions}
We developed and tested a prototype detector for the XAFS beamline at Elettra Synchrotron in Trieste based on the 
SDD technology.
The prototype demonstrated the importance of designing a modular detector to reach a larger collecting area while maintaining low pile-up and short dead time.
The prototype detector comprises a single SDD array of eight cells, each one coupled to a SIRIO ultra low noise preamplifier to form an independent channel.
The detector final version will be an $8 \times 8$ matrix for a total collecting area of about 576~mm$^2$, seven times larger than the detector currently employed at the beamline.
During preliminary tests, we measured the performance of the prototype detector in laboratory and obtained an energy resolution of 167~eV at 5.89 keV at $0^\circ$C.
We also performed a test beam on the prototype detector at the XAFS beamline of Elettra and verified its performance for high count rate. 

\ack
We acknowledge EUROFEL-MIUR and INFN R$\&$D ReDSoX projects for founding this activity.

\section*{References}

\bibliographystyle{iopart-num}
\bibliography{/home/sergio/Scrivania/Lavoro/GruppoSDD/bibliografia/bibliografia}  

\providecommand{\newblock}{}
\begin{thebibliography}{1}
\expandafter\ifx\csname url\endcsname\relax
  \def\url#1{{\tt #1}}\fi
\expandafter\ifx\csname urlprefix\endcsname\relax\def\urlprefix{URL }\fi
\providecommand{\eprint}[2][]{\url{#2}}

\bibitem{DiCicco2009}
{Di Cicco} A, {Aquilanti} G, {Minicucci} M, {Principi} E, {Novello} N,
  {Cognigni} A and {Olivi} L 2009 {\em Journal of Physics Conference Series\/}
  {\bf 190} 012043

\bibitem{Geant42010}
{Geant4 Collaboration} 2010 {Geant4: A Simulation Toolkit for the Passage of
  Particles through Matter} astrophysics Source Code Library (\textit{Preprint}
  \eprint{1010.079})

\bibitem{Bertuccio2014}
{Bertuccio} G, {Macera} D, {Graziani} C and {Ahangarianabhari} 2014 {\em
  Proceedings of the IEEE Nuclear Science Symposium Seattle – WA - USA, 8-15
  Nov.\/}

\bibitem{Bertuccio2015}
{Bertuccio} G, {Ahangarianabhari} M, {Graziani} C, {Macera} D, {Shi} Y,
  {Rachevski} A, {Rashevskaya} I, {Vacchi} A, {Zampa} G, {Zampa} N, {Bellutti}
  P, {Giacomini} G, {Picciotto} A and {Piemonte} C 2015 {\em Journal of
  Instrumentation\/} {\bf 10} P01002

\end{thebibliography}

\end{document}